\documentclass[showpacs,twocolumn]{revtex4}
\usepackage{amsfonts}
\usepackage{amsmath}
\usepackage{amssymb}
\usepackage{graphicx}

\begin{document}

\title{Photovoltaic transistor of atoms due to spin-orbit coupling in three optical traps}
\author{Haihu Cui$^{1}$, Mingzhu Zhang$^{2}$ and Wenxi Lai$^{2}$}\email{wxlai@pku.edu.cn}

\affiliation{1 Department of Building Engineering, Inner Mongolia Vocational and Technical College of Communications, Chifeng 024005, China}
\affiliation{2 School of Applied Science, Beijing Information Science and Technology University, Beijing 100192, China}

\begin{abstract}
In this paper, spin-orbit coupling induced photovoltaic effect of cold atoms has been studied in a three-trap system which is an two-dimensional extension of a two-trap system reported previously. It is proposed here that atom coherent length is one of the important influence to the resistance of this photovoltaic battery. Current properties of the system for different geometrical structures of the trapping potentials are discussed. Numerical results show extension in the number of traps could cause current increase directly. Quantum master equation at finite temperature is used to treat this opened system. This work may give a theoretical basis for further development of the photovoltaic effect of neutral atoms.
\end{abstract}

\maketitle

In the technique of atomtronics, atoms can be controlled and manipulated analogous to the operation in electronics~\cite{Seaman,Pepino,Ramanathan,Beeler,Eckel,Daley,Aghamalyan,Wilsmann,Ryu}. One kind of devices in this field is atomic battery. Until now, there are some models of this kind device, such as atomtronic battery based on asymmetric wells~\cite{Zozulya} and equivalent chemical potentials~\cite{Caliga}.

Quite recently, spin-orbit coupling induced photovoltaic effect has been proposed~\cite{wlai,Entin-Wohlman}. The photovoltaic system of atoms should be seen as another kind of atomic battery~\cite{wlai}. The spin-orbit coupling in cold atoms can be realized using two-photon Raman transition~\cite{Y-J-Lin} or clock transition combined with synthetic dimension of atoms~\cite{Mancini,Stuhl,Livi}. The clock transitions of alkaline-earth(-like) atoms are featured with long coherent times~\cite{Norcia}, which can be used for the study of photovoltaic battery of atoms~\cite{wlai}. In the clock transition induced spin-orbit coupling, two internal states of an atom could be coupled to the momentum of the atom. Then, atoms with different internal states move in different directions. As a result, atoms with the two different states would be collected on two sides of the system just like positive and negative charges collected on the two electrode of a electronic battery. The synthetic dimension of atoms consists of internal states of atoms and spatial dimensions~\cite{Boada,Celi}. It means at least two traps are needed in the configuration for the spin-orbit coupling. Therefore, the double-trap model of photovoltaic system proposed in the previous work~\cite{wlai} is the basic atomic component. It is given that current in the basic component is very limited.

In this paper, we extend the double-trap photovoltaic system to three-trap system for the exploration of its scalability in two dimensions. By adding a trap in the system, we plan to test properties of the photovoltaic system, such as the battery resistance due to limited atom coherent length and spatial structure of the optical potentials. In addition, we expect to obtain larger current than that in the original double-trap system. Quantum master equation of atom density matrix is used to describe the present opened system. In experiments, a few optical traps for bounding cold atoms can be manufactured~\cite{Caliga,Caliga2,Caliga3}, and they have potential applications for the study of this new kind of transistors.

\begin{figure}
\includegraphics[width=7cm]{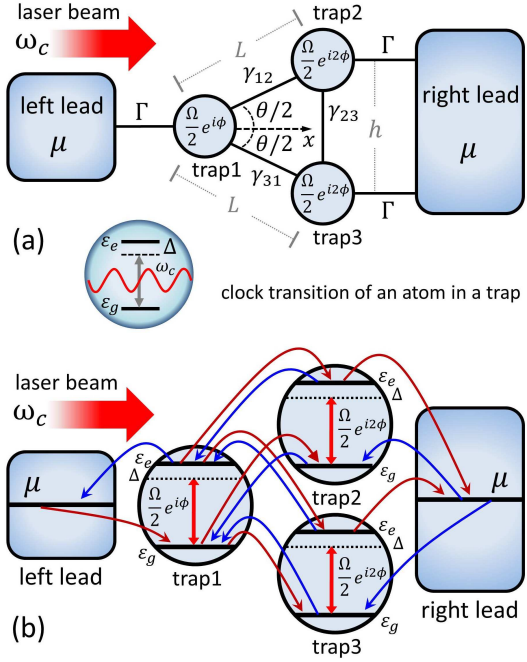}\
\caption{(Color on line) (a) Geometrical structure of the three-trap photovoltaic transistor. Three optical traps form a triangle with these traps located on three vortices, respectively. The distance between trap 1 and trap 2 is $L$, the distance between trap 1 and trap 3 is also $L$. $\theta$ is one angle of the isosceles triangle as shown in the figure. Each optical is coupled to a atomic bath. (b) Illustration of energy structure and atom movement in the photovoltaic transistor. Here, Chemical potentials $\mu$ in the left and right leads are the same and satisfy the relation $\varepsilon_{g}<\mu<\varepsilon_{e}$. Atoms can transport from one atomic bath to the other through the three-trap system. A coherent optical field would drive the clock transition of the atoms in the optical traps with Rabi frequency $\Omega$.}\label{sys}
\end{figure}

The geometrical structure of the model is conceptually shown in Fig.~\ref{sys} (a). The three optical traps and their inter-trap tunneling form a triangle with the three traps located on the three vertexes of the triangle. In this work, we just consider isosceles triangle for the arrangement of optical traps that the distance between trap 1 and trap 2 and the distance between trap 1 and trap 3 are the same, denoted by $L$. In this way, the distance $h$ between trap 2 and trap 3 would depend on the angle $\theta$ and the distance $L$. The direction of the laser beam is set to be always perpendicular to the connection line between trap 2 and trap 3.The angle between the direction of incident laser beam and the tunneling line between the trap 1 and trap 2 (or trap 3) is $\theta/2$. Here, we consider Fermion gas rare earth atoms $^{173}Yb$ which have clock transition between the ground state $g=$ $^{1}S_{0}$ and the metastable state $e=$ $^{3}P_{0}$. Wave length of the clock field driving this transition is around $\lambda_{C}=758$ nm~\cite{Barber,Gorshkov}. Just single atom occupation in a trap is considered for the convenience of theoretical calculations.

The three optical traps can be treated as quantum opened systems whose environment is the two atomic leads. The trap 1 is coupled to the left lead, and trap 2 and trap 3 are coupled to the right lead as illustrated in Fig.~\ref{sys}. The coupling between the traps and leads are all the same, characterized by the tunneling rate $\Gamma$. Atoms can transport from one atom bath to the other bath through the system of three optical traps. The chemical potentials in the two atom baths are set to be the same as $\mu$. It reveals the current through the battery here is not directly caused by the chemical potential bias like in Ref.~\cite{Caliga}. At temperature $T$, cold atoms in these baths can be described by the Fermi-Dirac distribution $f_{u}(\varepsilon_{s})=\frac{1}{e^{(\epsilon_{s}-\mu)/k_{B}T}+1}$, where $\varepsilon_{s}$ indicates atom internal energy with $s=g$ represents the ground state and $s=e$ denotes the excited state, and $k_{B}$ is the Boltzmann constant.

For the laser beam propagating in the horizontal direction, trap 1 is in one row, trap 2 and trap 3 are vertically arranging in another row with the same distance from the trap 1 as shown in Fig.~\ref{sys} (a). In this configuration, spin-orbit coupling can be occurred that when an atom is excited from its ground state to the excited state by a photon, at the same time, it acquires momentum from the photon and moves from one trap to the other. Strictly speaking, it is pseudo spin of atoms. The spin-orbit coupling could be reflected in atom-light coupling with a phase that related to momentum transfer between photon and atom. The phase can be seen as an effective magnetic flux of artificial gauge field~\cite{Mancini,Stuhl,Livi}. If an atom in trap 1 gets a phase $\phi_{1}=\phi$ due to the external beam action, the atoms in the trap 2 and trap 3 should get the relative phase $\phi_{2}=\phi_{3}=2 \phi$. The phase shift can be expressed as $\phi_{\alpha}=2\pi L \cos(\theta/2)/\lambda_{C}$ which is related to the the momentum change $2\pi cos(\theta/2)/\lambda_{C}$ of an atom along the direction $\theta/2$ and path length $L$ of tunneling between two traps. The net phase $\phi_{\alpha}$ can be seen as an artificial magnetic flux in the closed trajectory of atom transitions.

The whole Hamiltonian of the three-trap photovoltaic transistor can be written as ($\hbar=1$),
\begin{eqnarray}
H &=& \sum_{\alpha,s}\varepsilon_{s} a_{\alpha s}^{\dag}a_{\alpha s}+\sum_{\alpha\neq\beta,s}\gamma_{\alpha\beta}a_{\alpha s}^{\dag}a_{\beta s}+\sum_{u,k,s} \varepsilon_{k} b_{u ks}^{\dag}b_{u ks}\notag \\ &&+g \sum_{k,s} (b_{L ks}^{\dag}a_{1s}+b_{R ks}^{\dag}(a_{2s}+a_{3s})+H.c.)\notag \\ &&+\frac{\Omega}{2}\sum_{\alpha}(e^{i\omega_{c} t} e^{i \phi_{\alpha}} a_{\alpha g}^{\dag}a_{\alpha e}+h.c.).
\end{eqnarray}
In the first part of the Hamiltonian, $a_{\alpha s}$ ($a_{\alpha s}^{\dag}$) is annihilation (creation) operator of atoms in the trap at position $\alpha$ $(\alpha=1,2,3)$ and $s$ takes $g$ for the ground state, takes $e$ for the excited state. Corresponding energy levels are denoted as $\varepsilon_{g}$ and $\varepsilon_{e}$, respectively. The inter-trap coupling is characterized by the second part with the inter-trap tunneling rates $\gamma_{12}=\gamma_{31}=\gamma_{0}e^{-L/L_{0}}$ and $\gamma_{23}=\gamma_{0}e^{-h/L_{0}}$. The values of $\gamma_{\alpha\beta}$ for the indexes $\alpha$ and $\beta$ are symmetry, where $\alpha,\beta=1,2,3$. Then $\gamma_{0}$ can be understood as the tunneling rate when the distance between two traps is zero. $L_{0}$ represents atom coherent length in this optical lattice. The third part represents atom gas in the three baths, where $b_{u k s}$ ($b_{u k s}^{\dag}$) is annihilation (creation) operator of an atom in the left bath $u=L$ or the right bath $u=R$ with energy $\varepsilon_{k}$, wave number $k$, and internal state $s$. The tunnelings between the three-trap system and the atomic baths are described by the forth part of the Hamiltonian with tunneling amplitude $g$. In the final therm, atoms in the traps are coupled to the clock field with the Rabi frequency $\Omega$. The phase $\phi_{\alpha}$ is corresponding to the position $\alpha$ of vertex in the triangle structure.

Next, we use the quantum Liouville's equation $\partial\rho_{tot}/\partial t=-i[H,\rho_{tot}]$ to quantitatively describe the atom motion, where $\rho_{tot}$ is the total density matrix of the whole configuration. Using the free evolution Hamiltonian $H_{0}=\sum_{\alpha,s}\varepsilon_{s} a_{\alpha s}^{\dag}a_{\alpha s}+\sum_{u,k,s} \varepsilon_{k} b_{u ks}^{\dag}b_{u ks}$, the equation of motion can be transformed into the interaction picture,
\begin{eqnarray}
\frac{\partial\tilde{\rho}_{tot}}{\partial t}=-i[\tilde{H_{1}},\tilde{\rho}_{tot}],
\label{eq:int-equ}
\end{eqnarray}
where Hamiltonian $\tilde{H_{1}}$ in the commutator is
\begin{eqnarray}
\tilde{H_{1}}&=&\sum_{\alpha\neq\beta,s}\gamma_{\alpha\beta}a_{\alpha s}^{\dag}a_{\beta s}+\frac{\Omega}{2}\sum_{\alpha}(e^{-i(\Delta t-\phi_{\alpha})} a_{\alpha g}^{\dag}a_{\alpha e}+h.c.)\notag \\ &&+g \sum_{k,s} (b_{L ks}^{\dag}a_{1s}e^{i(\varepsilon_{k}-\varepsilon_{1s})t}+b_{R ks}^{\dag}(a_{2s}e^{i(\varepsilon_{k}-\varepsilon_{2s})t}\notag \\ &&+a_{3s}e^{i(\varepsilon_{k}-\varepsilon_{3s})t})+H.c.).
\label{eq:intH}
\end{eqnarray}
where $\Delta=\varepsilon_{e}-\varepsilon_{g}-\omega_{c}$ denotes the atom-light detunings being the same for all traps. Substituting the time integration of Eq. \eqref{eq:int-equ} into itself, one can reach
\begin{eqnarray}
\frac{\partial\tilde{\rho}_{tot}(t)}{\partial t}&=&-i[\tilde{H}_{2}(t)+\tilde{H}_{3}(t),\tilde{\rho}_{tot}(t)] \notag\\
&&-i[\tilde{H}_{4}(t),\tilde{\rho}_{tot}(0)] \notag\\
&&-\int_{0}^{t}[\tilde{H}_{4}(t),[\tilde{H_{1}}(t'),\tilde{\rho}_{tot}(t')]]dt'.
\label{eq:int-equ2}
\end{eqnarray}
where the simplified terms $\tilde{H}_{2}(t)=\sum_{\alpha\neq\beta,s}\gamma_{\alpha\beta}a_{\alpha s}^{\dag}a_{\beta s}$, $\tilde{H}_{3}(t)=\frac{\Omega}{2}\sum_{\alpha}(e^{-i(\Delta t-\phi_{\alpha})} a_{\alpha g}^{\dag}a_{\alpha e}+h.c.)$ and $\tilde{H}_{4}(t)=g \sum_{k,s} (b_{L ks}^{\dag}a_{1s}e^{i(\varepsilon_{k}-\varepsilon_{1s})t}+b_{R ks}^{\dag}(a_{2s}e^{i(\varepsilon_{k}-\varepsilon_{2s})t}+a_{3s}e^{i(\varepsilon_{k}-\varepsilon_{3s})t})+H.c.)$ in equation \eqref{eq:int-equ2} are just the three parts in equation \eqref{eq:intH}, respectively.

As large atomic reservoirs, left lead and right lead are assumed to be equilibrium atomic gas with corresponding time independent density matrices $\rho_{L}$ and $\rho_{R}$. Therefore, the total density matrix of the whole system could be written as $\rho_{tot}(t)=\rho(t)\rho_{L}\rho_{R}$, where $\rho$ is reduced density matrix of the double-trap system. Taking trace Tr over all microstates of the two leads, one can write the equation\eqref{eq:int-equ2} as,
\begin{eqnarray}
\frac{\partial\tilde{\rho}(t)}{\partial t}&=&-i[\tilde{H}_{2}(t)+\tilde{H}_{3}(t),\tilde{\rho}(t)] \notag\\
&&-\int_{0}^{t}Tr[\tilde{H}_{4}(t),[\tilde{H}_{4}(t'),\tilde{\rho}(t')\rho_{L}\rho_{R}]] dt',
\label{eq:int-equ3}
\end{eqnarray}
where the reduced density matrix is $\tilde{\rho}(t)=Tr[\tilde{\rho}_{tot}(t)]$. In the second term of the right side of Eq. \eqref{eq:int-equ3}, the trace would actually give rise to the Fermi-Dirac distribution function $Tr[b_{u k s}^{\dag}b_{u k s}\rho_{u}]=f_{u}(\varepsilon_{k})$. In Born-Markov approximation, $\tilde{\rho}(t')$ in Eq. \eqref{eq:int-equ3} can be written into $\tilde{\rho}(t)$~\cite{Scully}, which simply the integration over time $t'$. Using unitary transformation $e^{-it\sum_{\alpha}\Delta a_{\alpha e}^{\dag}a_{\alpha e}}$ to transform Eq. \eqref{eq:int-equ3} back to Schr\"{o}inger picture, one could achieve the equation which describes evolution of the three-trap opened system~\cite{Scully,W-Lai2,W-Lai3},
\begin{eqnarray}
\frac{\partial\rho}{\partial t} &=&-i [H_{sys},\rho]+\sum_{\alpha=1}^{3}\mathcal{L}_{\alpha}\rho,
\label{equation}\end{eqnarray}
where $\rho(t)=e^{-it\sum_{\alpha}\Delta a_{\alpha e}^{\dag}a_{\alpha e}}\tilde{\rho}(t)e^{it\sum_{\alpha}\Delta a_{\alpha e}^{\dag}a_{\alpha e}}$. The first term on the right side of Eq.\eqref{equation} represents free evolution of atom transitions in the synthetic dimension of three traps. The effective Hamiltonian of this free evolution is written in the time independent form
\begin{eqnarray}
H_{sys}&=&\sum_{\alpha}\Delta a_{\alpha e}^{\dag}a_{\alpha e}+\sum_{\alpha\neq\beta,s}\gamma_{\alpha\beta}a_{\alpha s}^{\dag}a_{\beta s} \notag\\
&&+\frac{\Omega}{2}\sum_{\alpha}(e^{i \phi_{\alpha}} a_{\alpha g}^{\dag}a_{\alpha e}+H.c.),
\label{effH}\end{eqnarray}
where $\Delta=\varepsilon_{e}-\varepsilon_{g}-\hbar \omega_{c}$ describes the detuning between the clock field frequency and the two transition levels $\varepsilon_{g}$ and $\varepsilon_{e}$ of an atom.

Coupling between the three optical traps and three atomic baths is described by the incoherent term $\sum_{\alpha}\mathcal{L}_{\alpha}\rho$, considering the Born-Markov approximation in Eq.\eqref{equation}. The Liouville super-operators $\mathcal{L}_{L}$ and $\mathcal{L}_{R}$ acting on the density matrix $\rho$ can be written as
\begin{eqnarray}
\mathcal{L}_{\alpha}\rho&=&\frac{\Gamma}{2}\sum_{\alpha,s}[f_{\alpha}(\varepsilon_{s})(2a_{\alpha,s}^{\dag}\rho a_{\alpha,s}-\{a_{\alpha,s}a_{\alpha,s}^{\dag},\rho\})\notag\\
&&+(1-f_{\alpha}(\varepsilon_{s}))(2a_{\alpha,s}\rho a_{\alpha,s}^{\dag}-\{a_{\alpha,s}^{\dag}a_{\alpha,s},\rho\})],
\end{eqnarray}
with the anti-commutation relation $\{O,\rho\}$ for any operator $O$. The coupling strength $\Gamma$ in detail is $\Gamma=2\pi|g|^{2}D(\varepsilon_{s})$, where $D(\varepsilon_{s})$ is the density of states of atoms in the lead at energy $\varepsilon_{s}$. For an atom with definite discrete state in the trap, its state distribution is much narrower than the atom state distribution in the lead. As a result, atom in the trap would feel that density of states of in the lead is almost a constant. Therefore, $\Gamma$ will be taken as a constant in the numerical treatment here.

According to the atom number conservation, difference of left current $I_{L}$ and right current $I_{R}$ at a time $t$ should be equal to the rate of atom number change in the three-trap system~\cite{Davies,Jauho,Twamley},
\begin{eqnarray}
\frac{d}{d t}\sum_{\alpha=1}^{3} (\langle n_{\alpha}\rangle)=I_{L}-I_{R},
\label{current}\end{eqnarray}
where the mean value of atom number is $\langle n_{\alpha}\rangle=\langle\sum_{s}a_{\alpha s}^{\dag}a_{\alpha s}\rho\rangle$ for $\alpha=1,2,3$ with $\langle\rangle$ represents quantum average over all state of the system. One can substitute Eq.\eqref{equation} into Eq.\eqref{current} and obtain detail expressions of current as
\begin{eqnarray}
I_{L}=\Gamma \sum_{s}(f_{L}(\varepsilon_{s})\langle a^{\dag}_{1 s}\rho a_{1 s}\rangle-(1-f_{L}(\varepsilon_{s}))\langle\rho a^{\dag}_{1 s} a_{1 s}\rangle),
\label{eq:lefti}
\end{eqnarray}
and
\begin{eqnarray}
I_{R}=\Gamma\sum_{s} ((1-f_{R}(\varepsilon_{s}))\langle\rho a^{\dag}_{2 s} a_{2 s}\rangle-f_{R}(\varepsilon_{s})\langle a^{\dag}_{2 s}\rho a_{2 s}\rangle) \notag \\
+\Gamma \sum_{s}((1-f_{R}(\varepsilon_{s}))\langle\rho a^{\dag}_{3 s} a_{3 s}\rangle-f_{R}(\varepsilon_{s})\langle a^{\dag}_{3 s}\rho a_{3 s}\rangle).
\label{eq:righti}
\end{eqnarray}
The left current $I_{L}$ is related to trap 1 as shown in Fig.~\ref{sys}, therefore it is proportional to atom number distribution in trap 1, namely, $Q_{1}=\sum_{s}(f_{L}(\varepsilon_{s})\langle a^{\dag}_{1 s}\rho a_{1 s}\rangle-(1-f_{L}(\varepsilon_{s}))\langle\rho a^{\dag}_{1 s} a_{1 s}\rangle)$ and inversely proportional to $\tau=\Gamma^{-1}$. $Q$ and $\tau$ can be seen as effective charge and atom transit rate. Therefore, the current satisfy the definition $I_{L}=Q_{1}/\tau$ conceptually. In the same way, the right current $I_{R}$ is depends on atom number distribution in trap 2 and trap 3. As a result, $I_{R}$ is proportional to the effective charges $Q_{2}=\sum_{s} ((1-f_{R}(\varepsilon_{s}))\langle\rho a^{\dag}_{2 s} a_{2 s}\rangle-f_{R}(\varepsilon_{s})\langle a^{\dag}_{2 s}\rho a_{2 s}\rangle)$ and  $Q_{3}=\sum_{s}((1-f_{R}(\varepsilon_{s}))\langle\rho a^{\dag}_{3 s} a_{3 s}\rangle-f_{R}(\varepsilon_{s})\langle a^{\dag}_{3 s}\rho a_{3 s}\rangle)$, respectively. In the same way, we have $I_{R}=(Q_{2}+Q_{3})/\tau$. Here, we have defined the direction of current from the left bath to the right bath is positive. Therefore, the total current can be written as $I=(I_{L}+I_{R})/2$, in the form of average value. In Eqs.\eqref{eq:lefti} and \eqref{eq:righti} of current, probabilities of atoms in the three traps are involved, they are probability of empty trap $\alpha$, $P_{\alpha 0}=\langle a_{\alpha s}a^{\dag}_{\alpha s}\rho \rangle$, probability of the ground state atom in trap $\alpha$, $P_{\alpha g}=\langle\rho a^{\dag}_{\alpha g} a_{\alpha g}\rangle$, and probability of the excited state atom in trap $\alpha$, $P_{\alpha e}=\langle\rho a^{\dag}_{\alpha e} a_{\alpha e}\rangle$, where $\alpha=1,2,3$.

\begin{figure}
\includegraphics[width=8cm]{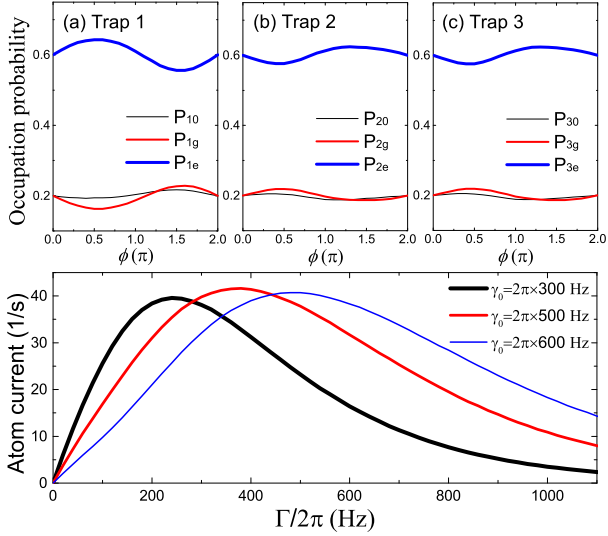}\
\caption{(Color on line) (a) Atom distribution probability in trap 1. (b) Atom distribution probability in trap 2. (c) Atom distribution probability in trap 3. In the (a)-(c) three figures, $\Omega=2\pi\times 600$ Hz. (d) Current versus trap-lead transit rate $\Gamma$ for different trap-trap tunneling rate $\gamma_{0}$. The rest corresponding parameters are $\theta=\pi/3$, $L=1000$ nm, $\triangle=\Omega/2$.}\label{igama}
\end{figure}

In Fig.~\ref{sys} (b), energy levels and trajectories of atom motion have been illustrated. The three traps are coupled each other through corresponding state ($e\leftrightarrow e$, $g\leftrightarrow g$) atom tunnelings. Chemical potentials in the left and right leads are the same as $\mu$. The chemical potential satisfies the energy configuration of $\varepsilon_{g}<\mu<\varepsilon_{e}$. It reveals that energy of atoms in the leads would be higher than the ground levels and lower than the excited levels in the three traps. Therefore, atoms in the leads can directly transfer into the ground states of the three traps. However, if one wants to send atoms from the three traps to any of the two leads, atoms must be in their excited states. The applied field with clock transition frequency $\omega_{c}$ is set to excite atoms in the traps with the Rabi frequency $\Omega$. The different distribution of occupation probabilities in the three traps is created by the relative phase $\phi_{2}-\phi_{1}=\phi$ ($\phi_{2}=\phi_{3}$). Due to the phase of an atom in trap 1 is different from the phase of an atom in trap 2 ( or trap 3 ), the occupation probability of excited atom in trap 1 would be different from that of excited atom in trap 2 ( or trap 3 ) as shown in Fig.\ref{igama} (a)-(c). As illustrated in Fig.\ref{igama} and following results, the different distribution of occupation probabilities of the optical traps induces photovoltaic effect and give rises to net current between the two leads. The relative phase $\phi$ represents artificial gauge field induced by the spin-orbit coupling of atom-light interaction. In other words, an excited atom gains a momentum $cos(\theta/2)/\lambda_{C}$ from a photon and transits from one trap to the other trap, which causes coupling between pseudo spin and momentum of the atom. Therefore, phase of an atom in one trap is different from phase of an atom in the other trap. The phase difference of atoms in the left and right traps is involved in the atom-light coupling Hamiltonian, which leads to population difference of atom occupations in the left and right traps. Since the rate of particle transport in a transistor depends on the feature of particle occupations, the phase difference finally induce net photocurrent in the system.

\begin{figure}
\includegraphics[width=7cm]{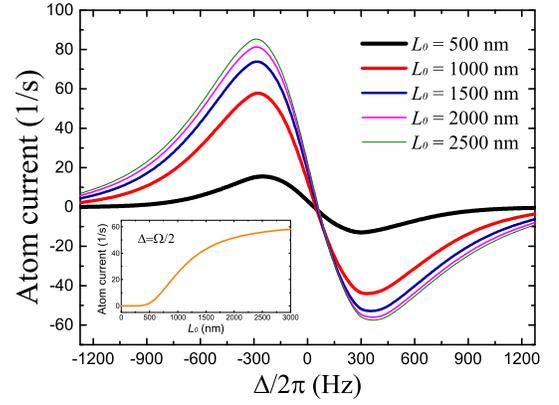}\
\caption{(Color on line) Atomic current as a function of atom-light detuning for different atom coherent length. The corresponding parameters are $\theta=\pi/3$, $L=1000$ nm, $\Omega=2\pi\times 600$ Hz. Inset: Current versus the coherent length $L_{0}$ for a fixed detuning $\triangle=\Omega/2$.}\label{idelta}
\end{figure}

The rates of trap-lead transit $\Gamma$ and inter-trap tunneling $\gamma_{0}$ determine characteristic time of atom movement in the transistor. Too slow motion of atom decreases atomic current. However, too fast atom transfer make the transistor hard to polarize atom population distribution in the traps, which also depresses stationary net current of atoms. Therefore, there would be current peak for the change of the transit rate $\Gamma$ as plotted in Fi.g \ref{igama} (d). The figure also show that the current peak depends on the inter-trap tunneling rate $\gamma_{0}$. For smaller $\gamma_{0}$, the top current is positioned at lower $\Gamma$, in contrast, for larger $\gamma_{0}$, the top current is landed at the value of higher $\Gamma$. It reveals that to achieve largest current, matching between the rates of trap-lead transit and inter-trap tunneling is very important.

The distance, $L$ and $h$, between two traps is about several hundreds nanometers. At the same time, order of magnitudes of atom coherent length should be comparable to the scale of optical traps. How the change of coherent length of atoms affects the resistance of the photovoltaic battery ? To answer this question, we plot atom current as a function of atom-light detuning $\Delta$ for different atom coherent lengths in Fig.~\ref{idelta}. When coherent length of atoms very short, for example, $L_{0}<L$, atom current is weak, lower than 20 $s^{-1}$. When the coherent length is around the trap-trap distance $L$, the current would increase remarkably, close to 60 $s^{-1}$. For further increase of this coherent length, larger current can be seen from the Fig.~\ref{idelta}. Increase in current is become slower for the same scale of $L_{0}$ increase as shown in the inset figure. The variation of current comes from the rate $\gamma_{\alpha\beta}$ of atom tunneling which is related to the coherent length $L_{0}$. Equivalently, the ratio of trap-trap distance and atom coherent length $L/L_{0}$ influences the resistance in the battery. The basic parameters used in Fig.~\ref{idelta} and the following figures are $\gamma_{0}=2\pi\times 500$ Hz, $\Gamma=2\pi\times 400$ Hz and $k_{B}T=0.1\Gamma$\cite{Livi}.

\begin{figure}
\includegraphics[width=7cm]{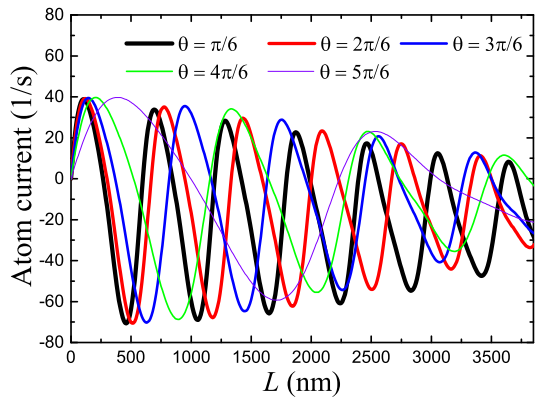}\
\caption{(Color on line) Atomic current as a function of on trap-trap distance $L$ under different angles $\theta$ of trap arrangement. The corresponding parameters are chosen as $L_{0}=5$ $\mu m$, $\Omega=2\pi\times600$ Hz, $\Delta=2\pi\times300$ Hz.}\label{itheta}
\end{figure}

Geometry of the trapping potentials can be controlled in experiments~\cite{Petsas,Gullans}. The fact allow us to test the properties of this kind of battery for different geometrical structure of the trap arrangements. Fig.~\ref{itheta} shows the current of the system obviously depends on angle $\theta$ and distance $L$ of the traps. When the angle $\theta$ is small, current variation for the change of trap distance $L$ fluctuates frequently. On the contrary, when the angle $\theta$ is close to $\pi$, the current fluctuation becomes remarkably slow. From the other point of view, the above characteristics of the system also reveals geometrical structure of optical potentials may be probed using the photovoltaic effect. Current amplitude fluctuates for the change of trap-trap distance $L$, because current depends on the phase $\phi$ and the phase is proportional to $L cos(\theta/2)$ ( projection of $L$ along the horizontal coordinate $x$ ). It means $cos(\theta/2)$ is the proportionality coefficient between $\phi$ and $L$. For small angle $\theta$, the coefficient $cos(\theta/2)$ is very large, as a result, $\phi$ increase quickly for the change of $L$. Therefore, the current fluctuation is fast in Fig.~\ref{itheta}. On the contrary, when $\theta$ is close to $\pi$, the proportionality coefficient $cos(\theta/2)$ tends to zero. It leads to the slow change of $\phi$ and furthermore slow fluctuation of corresponding current.

\begin{figure}
\includegraphics[width=7cm]{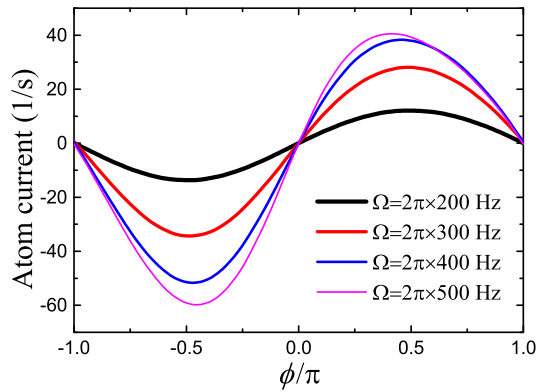}\
\caption{(Color on line) Atomic current vs artificial gauge phase $\phi$ for different Rabi frequency $\Omega$ under the parameters $\Delta=2\pi\times 220$, $L_{0}=10$ $\mu m$, $L=500$  nm, $\theta=\pi/4$.}\label{iphi}
\end{figure}

The Josephson effect phase $\phi$ depend current behavior in the system of three optical traps is similar to the result that obtained in the two optical traps\cite{wlai}, as illustrated in Fig.~\ref{iphi}. Calculations in both Fig.~\ref{iphi} and Fig.~\ref{idelta} show that current in the three-trap battery is remarkably larger than the current in the two-trap battery at the same parameters. Therefore, extensions of the system in trap number can increase current. Large current could be speculated if an optical lattice with a great number of trapping potential is used in the photovoltaic system. The fact is important for future applications of the present model. Value of the current is under complete control in principle through the parameters of the system, such as Rabi frequency, distance and depth of trapping potential, number of traps.

Experimental realization of the photovoltaic battery with a few trapping potentials is feasible, as atom transistor has been tested preliminarily nowadays~\cite{Caliga,Caliga2,Caliga3,Mancini,Stuhl,Livi}. The challenges may come when one try to construct such system in a optical lattice with many trapping potentials. Since, on one hand, a large array of optical traps would be needed to couple with two deep potential of atom gas; on the other hand, in this configuration, atoms are cooled enough to ensure coherent length of atoms are closer or longer than the scale of potential period. Since the atom-light transition frequency is in the range of visible light frequency, room temperature $T$ can not effectively influence the atom transition here. However, atoms should be at low temperature, as they would be controlled in the optical potential and coherent tunneling should be allowed. Therefore, cold atoms should be used to implement this process in experiment at low enough temperature unless deep enough potential for the control of atoms can be created in practice.

In Summery, it is probed that the basic component of the photovoltaic transistor is scalable, in which current increase has been observed in theoretical calculations by just adding the number of trapping potentials. There is no asymmetric wells and no chemical potential difference in the transistor, atom current in the leads is obtained due to the clock transition induced spin-orbit coupling in synthetic dimension. The resistance of the photovoltaic battery could be come from the coherent length of atoms. Geometrical structure of optical lattice also affect the current behavior. Even though the model is based on single atom tunneling process, the result should be appropriate to the case of noninteracting many atoms bounded in traps.

\begin{acknowledgments}
This work was supported by the Scientific Research Project of Beijing Municipal Education Commission (BMEC) under Grant No. KM202011232017.
\end{acknowledgments}

\end{document}